# Ontology-based industrial data management platform

Sergey Gorshkov[1] [0000-0002-0821-8050],
Alexander Grebeshkov[1] [0000-0001-8899-414X], Roman Shebalov[1]

[1] TriniData LLC, 40-21 Mashinnaya str., 620089, Ekaterinburg, Russia
`serge@trinidata.ru`

**Abstract.** Relational and noSQL storages are developed for the fast processing of the large data sets having a stable structure, while the ontologies are used to represent complex and dynamic sets of information of a limited size. In the industrial applications it is often needed to maintain the large warehouses of data consolidated from various sources. The ontologies are useful to represent the structure of that data, but RDF triple stores are not well suitable for storing it. We offer an approach and a system allowing to use the opportunities of fast storage engines along with the flexibility of ontology-based data management tools, including SPARQL queries. The system implements a multi-model data abstraction layer which allows working with the data as if it is situated in RDF triple store, executes SPARQL queries over it and applies SHACL constraints and rules.

## 1    Introduction

Ontologies are quite useful for representing knowledge of a complex and changing structure. They minimize the gap between a conceptual representations and data structures. However, when representing industrial data using ontologies, we find that the conceptual part of ontology, TBox, is much more compact than its factual part, ABox.

The RDF triple stores are a class of storages designed especially for ontologies. They store TBox and ABox in a uniform way. This allows to query ontologies with SPARQL and reason over it, but also disallows some techniques which might be used to speed up data access: creating data-specific indexes in relational and noSQL storages, data sets segmentation, etc. This inspires the idea of storing the ABox data in the fast and scalable relational or noSQL database, providing a SPARQL interface for querying it.

A typical industrial use case for ontologies is an enterprise Knowledge graph, a unified representation of a wide amount of corporate data consolidated from various sources. Two main options of its implementation are a) building a physical data warehouse by importing all the data into a single storage (this differs from the "data lake" concept, because data is consolidated and transformed according to the common schema), and b) creating a "logical data mart", which extracts data from the external sources at the query execution.

To meet the industrial demand of an ontology-based data platform suitable for enterprise knowledge graph implementation, we have developed the ArchiGraph data virtualization platform, which meets the following key functional requirements:
1) Data structure (TBox) is managed with ontology-based tools and stored in the RDF triple store.
2) Materialized factual data (ABox) may be assigned to various storages according to the class(es) to which they belong. The platform shall have a management tool allowing to plug in new storages at run time and distribute data between them. Individuals of each class may be assigned to several storages.
3) Individuals of some classes may be associated with the external storages and extracted on demand according to the "logical data mart" paradigm.
4) All the data is available through a set of APIs, including SPARQL.
5) The platform supports applying the SHACL Constraints and SHACL Rules to the data stored in all the plugged storages.

The platform includes a data abstraction layer, which allows to use any suitable third-party database or their combination as a single multi-model ontology-based data storage, along with the rules execution engine, which can execute SPARQL queries over the data in these storages. This allow to combine advantages and performance of specialized data storages with the power and flexibility of ontology-based tools.

## 2 Related work

The need of heterogenous data integration in industrial projects using Semantic web technologies is widely discussed. The role of ontologies for representing the common data schema and mapping rules is acknowledged in several researches. However, their authors pay more attention to the data representation and transformation than to its storing and processing. An example of this approach can be found in (Akinyemi, 2018), where the integrated data are represented as an RDF file. This imposes limitations on the data size and access speed, even if the data are placed in the RDF triple store or split between several files or stores. Also, any materialized set of unified data can shortly become unactual due to the changes of the source data.

The OBDA paradigm offers direct access to the source databases through the semantic layer performing queries and data translation using mapping rules from source data structures to the common ontology. This approach is well studied and described, particularly, in (Calvanese, 2017; Lanti, 2017) for RDBMS and in (Cure, 2012; Araujo, 2017) for NoSQL storages. This approach can process larger data sets and always provides actual information, but reasoning implementation is problematic with it.

Another way of achieving the same goal is to extend applicability of RDF triple stores for processing large data amounts with clustering and other techniques. The overview of the possible implementations is given in (Pivert, 2018). However, we suppose that these are limited in performance due to the complex nature of the solutions.

The task of the corporate data consolidation into a single knowledge graph and the system architectures allowing to achieve this are discussed for example in (Wauer, 2010) and the works presenting the result of the Optique project (Kharlamov, 2017).

## 3      Platform architecture

The main motivation of our platform implementation is to let the data reside in the store that has the best performance and features according to the data nature – or in the storage where it originates. For example, MongoDB has a good functionality for handling geospatial data in GeoJSON format, Elasticsearch is the powerful full-text search engine, relational databases (particularly Postgres) are good for time series handling, HBase has the built-in support of timestamped attribute values and allows distributed data processing. There are attempts to implant these features into triple stores, such as SPARQL geospatial extensions or Lucene connectors, but we believe that for the better performance and faster integration with the emerging storages our way is more suitable. The architecture we have implemented consists of the components shown on the Fig. 1.

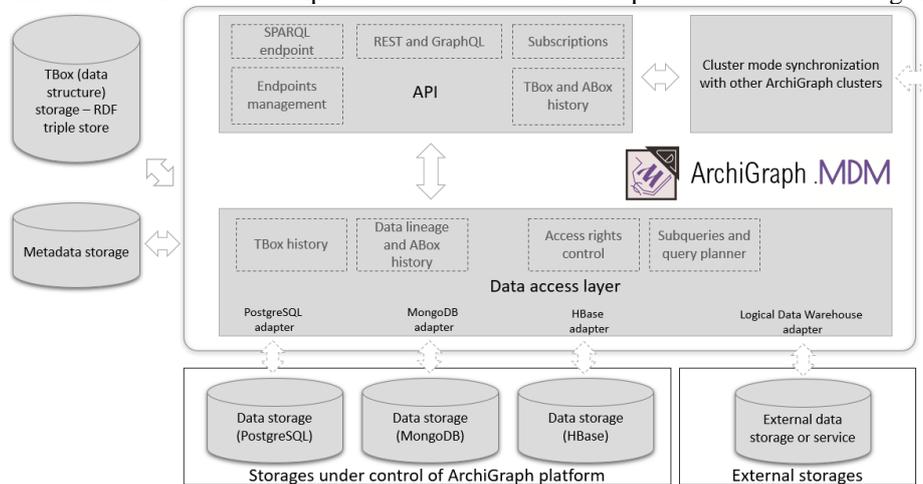

Fig. 1. Ontology-based data management platform architecture

The upper layer, Data access and management API, provides several interfaces for interacting with the model and data, including REST and GraphQL services with the simplified request formats along with the traditional SPARQL endpoint interface[1]. The requests are processed by the Data access layer which addresses the storages where the individuals of particular classes reside. It implements the queries dispatching logic required to process requests and builds execution plans. In this architecture the TBox (classes and properties definitions) is situated in the real RDF triple store, while the most part of the ABox (individuals) is distributed between various storages. The Storage adapters implemented at this layer provide a unified way of working with the ABox objects and take care of the optimal internal data structure in the storage, indexing and using specific storage functionality for achieving a better performance. Each class of the model may be assigned to one or more storages where its individuals are stored. The individuals of the unassigned classes remain in the triple store. If an individual belongs to more than one class, it can be replicated to several stores.

---

[1] The platform demo and documentation, including API reference, are available at https://archi-graph.pro

The proposed architecture is probably not the best choice for executing complex queries traversing the relations chains between the objects of different classes, but it is quite good for fast querying large sets of individuals, even with subqueries.

It is important that each storage may be clustered/sharded using its native features or the tools of its ecosystem, so the data storage may be distributed between many VMs. The ArchiGraph platform itself, in turn, may be clustered: a) the core components of the platform are run in the pods of a Kubernetes cluster, b) each instance of ArchiGraph may serve as a storage for another instance, c) several instances may synchronize data between them by subscribing on data updates. This allows nearly unlimited scaling and load balancing in the production environments.

## 4   Evaluation

We have performed tests to compare performance of our platform with the pure RDF triple store (Apache Fuseki). We have generated the uniform data set containing three classes (Company, Person and Project), whose individuals are interlinked: a Person can work in a Company, a Company can participate in a Project, a Person can be responsible for a Project. We have populated it with 1 000 000 Companies, the same number of Persons and x3 number of Projects (5 000 000 objects in total). All the individuals were stored in three copies: the first one in Fuseki, the second one in the Postgres database, and the third one in another Postgres database on the same machine.

Then we have run tests to assess data read performance in five scenarios: 1) pure Fuseki, 2) single Postgres database through ArchiGraph REST API, 3) single Postgres database through ArchiGraph SPARQL endpoint, 4) single Postgres database through ArchiGraph in the Logical Data Mart mode, 5) two Postgres databases (with data consolidation) through ArchiGraph in the Logical Data Mart mode.

Fuseki performs fast the queries involving graph traversal, such as "find all the projects of company X in which Y is a responsible person". Such queries take roughly 0,09 sec on Fuseki on our test virtual machine, comparing with 0,30 sec with ArchiGraph SPARQL endpoint and 0,28 sec with ArchiGraph REST service. In such cases ArchiGraph is slower due to the data consistency checks, and the fact that it extracts all the values of the properties of each extracted object and finds rdfs:label of each referenced object (this is done to simplify data display in the client applications). But there are several kinds of requests, rather frequent in the industrial applications, in which ArchiGraph is much more efficient than Fuseki – see Table 1 (time is averaged over 10 tests). The numbers in the column headers correspond to the test scenarios listed above.

Table 1. Query performance comparison, average execution time in seconds

| Query | Fuseki (1) | REST (2) | SPARQL (3) | LDM (4) | LDM (5) |
|---|---|---|---|---|---|
| Filter by text (1) | 16,43 | 1,09 | 1,10 | 2,18 | 1,06 |
| Filter by date (2) | 7,56 | 0,50 | 0,74 | 0,88 | 0,52 |
| Offset search (3) | 10,24 | 0,93 | 1,40 | - | - |

The queries are:
1) SELECT * WHERE { ?org rdf:type :Organization. ?org rdfs:label ?name FILTER(CONTAINS(STR(?name), "some text")) }

2) SELECT * WHERE { ?person rdf:type :person. ?person :birthDate ?birth. ?person rdfs:label ?name.  FILTER( ?birth > "1999-12-27"^^xsd:date ) } LIMIT 10
3) SELECT * WHERE { ?project rdf:type :Project. ?project rdfs:label ?name } OFFSET 990000 LIMIT 1000

Higher efficiency is achieved due to indexes on Postgres tables used by ArchiGraph.

## 5    Conclusions

We have demonstrated an architecture of the ontology-based data management platform, which provides transparent access to the data dispersed between several storages, using SPARQL query interface. The ArchiGraph platform industrial applications have demonstrated a good performance, flexibility and scalability of the solution. One of its valuable features is the possibility of changing the set of physical data storages in the runtime and redistributing data between them.

The query execution engine allows SHACL rules execution and constraints checking, which is useful in various scenarios such as data flow processing (including IoT use cases), near-real time complex logic calculations for decision-making support.

Although the proposed approaches are rather specific, they maintain interoperability with the Semantic Web specifications and tools. It is important in the context of building a corporate Knowledge graph, which may import several ontologies and use various ontology-based data transformation techniques.